\documentclass[sigconf]{acmart}
\usepackage{tabularx}
\usepackage{booktabs}
\usepackage{array}
\newcolumntype{Y}{>{\raggedright\arraybackslash}X}

\AtBeginDocument{%
  }

\setcopyright{acmlicensed}
\copyrightyear{2026}
\acmYear{2026}
\acmDOI{XXXXXXX.XXXXXXX}
\acmConference[EASE '26]{International Conference on Evaluation and Assessment in Software Engineering}{June 09--12,
  2026}{Glasgow, UK}
\acmISBN{978-1-4503-XXXX-X/2026/06}

\begin{document}

\title{ Evaluation of LLM-Based Software Engineering Tools: Practices, Challenges, and Future Directions}


\author{Utku Boran Torun}
\email{boran.torun@bilkent.edu.tr}
\author{Veli Karakaya}
\email{veli.karakaya@ug.bilkent.edu.tr}
\affiliation{%
  \institution{Bilkent University}
  \city{Ankara}
  \country{Turkey}
}

\author{Ali Babar}
\email{ali.babar@adelaide.edu.au}
\affiliation{%
  \institution{Adelaide University}
  \city{Adelaide}
  \country{Australia}
}

\author{Eray Tüzün}
\email{eraytuzun@cs.bilkent.edu.tr}
\affiliation{%
  \institution{Bilkent University}
  \city{Ankara}
  \country{Turkey}
}

\renewcommand{\shortauthors}{Torun et al.}

\begin{abstract}
Large Language Models (LLMs) are increasingly embedded in software engineering (SE) tools, powering applications such as code generation, automated code review, and bug triage. As these LLM-based AI for Software Engineering (AI4SE) systems transition from experimental prototypes to widely deployed tools, the question of what it means to evaluate their behavior reliably has become both critical and unanswered. Unlike traditional SE or machine learning systems, LLM-based tools often produce open-ended, natural language outputs, admit multiple valid answers, and exhibit non-deterministic behavior across runs. These characteristics fundamentally challenge long-standing evaluation assumptions such as the existence of a single ground truth, deterministic outputs, and objective correctness. In this paper, we examine LLM evaluation as a general, task-dependent concept through the lens of SE tasks. We discuss why reliable evaluation is essential for trust, adoption, and meaningful assessment of LLM-based tools, summarize the current state of evaluation practices, and highlight their limitations in realistic AI4SE settings. We then identify key challenges facing current approaches, including the absence of stable ground truth, subjectivity and multi-dimensional quality, evaluation instability due to non-determinism, limitations of automated and model-based evaluation, and fragmentation of evaluation practices. Finally, we outline future directions aimed at advancing LLM evaluation toward more robust, scalable, and trustworthy methodologies, to stimulate discussion on principled evaluation practices that can keep pace with the growing role of LLMs in SE.
\end{abstract}




%\received{2 March 2026}

\maketitle

\section{Introduction}
Large Language Models (LLMs) have rapidly become a central component of modern software engineering (SE) tools. They now power a wide range of AI4SE applications, including code generation, automated code review, bug triage, test generation, and developer assistance \cite{fan2023large}. Unlike earlier generations of software engineering tools, such as static analysis and rule-based systems like Facebook Infer\footnote{https://fbinfer.com/docs/man-infer/} and SonarQube\footnote{https://www.sonarsource.com/products/sonarqube/}, which were typically designed for narrowly scoped tasks, LLM-based systems function as general-purpose components that can be adapted to a wide range of software engineering activities through prompting and contextualization.. As a result, LLMs are increasingly embedded directly into developers’ everyday workflows and decision-making processes.

This growing adoption raises a fundamental question: how should LLM-based SE tools be evaluated? Evaluation plays a critical role in determining trust, guiding deployment decisions, and enabling meaningful comparison between tools and models. However, evaluating LLM-based systems poses challenges that differ substantially from those encountered in traditional SE or machine learning settings. Many LLM-driven software engineering tasks involve interpreting and generating context-dependent natural language artifacts, where quality cannot be assessed through a single predefined correct outcome. Moreover, LLMs are inherently stochastic, often producing different outputs across repeated runs under identical conditions \cite{ouyang2025empirical, atil-etal-2025-non}.

Existing evaluation practices struggle to fully account for these characteristics. While benchmarks such as SWE-Bench\footnote{https://www.swebench.com/}, Live Code Bench\footnote{https://livecodebench.github.io/index.html}, and automatic metrics have long been used to assess model performance, they often rely on assumptions such as deterministic behavior, fixed ground truth, or surface-level similarity between outputs and references \cite{papineni2002bleu,lin2004rouge}. These assumptions break down in many AI4SE scenarios, particularly for tasks such as code review and code explanation, where quality is subjective, and correctness is not binary. As a result, current evaluation methods may provide incomplete, unstable, or misleading signals about real-world performance. In response to these challenges, a diverse ecosystem of LLM evaluation approaches has emerged, including manual human evaluation, standardized benchmarks, reference-based automatic metrics, LLM-as-a-judge techniques, and comprehensive evaluation frameworks such as DeepEval\footnote{https://deepeval.com}, RAGAS\footnote{https://docs.ragas.io/en/stable/}, and HELM\footnote{https://crfm.stanford.edu/helm/}. While these approaches offers valuable insights, they also introduce new trade-offs related to scalability, reproducibility, bias, and alignment with developer needs. The lack of a unified understanding of their strengths and limitations has led to fragmentation in evaluation practice and uncertainty about how results should be interpreted.
%OVERVIEW OF PAPER

%In this study, we take a step back and examine the evaluation of LLM-based SE tools from an SE perspective. Rather than proposing a new evaluation framework or metric, our goal is to clarify why evaluation has become a central challenge for LLM-based AI4SE systems, summarize the current state of evaluation practice, and articulate the key challenges and open research questions that remain unresolved. By synthesizing insights from existing tools, benchmarks, and empirical observations, we aim to provide a structured view of where the field stands today and outline directions for future research on principled and trustworthy LLM evaluation in SE.

\section{Why Is LLM Evaluation Important?}
Evaluation determines whether LLM-based software engineering tools are useful, reliable, and safe in practice. In AI4SE, it serves as quality assurance for probabilistic systems whose behavior varies across prompts, contexts, and model versions. Without principled evaluation, tools may be optimized for the wrong objectives and fail to transfer to real-world workflows. In the following, we outline the key reasons why evaluation is central to LLM-based SE tools.

\textbf{Trust, Adoption, and Decision-Making in Practice.}
LLMs are now embedded in a wide range of SE activities, including code review, bug analysis, and documentation. In these settings, model outputs are often advisory rather than executable, meaning that developers must decide when to trust, accept, or ignore model suggestions \cite{barke2023grounded}. Evaluation, therefore, directly influences user trust and long-term adoption.
Without a reliable evaluation, it becomes difficult to assess whether an LLM accurately and consistently provides helpful guidance or merely produces plausible-looking responses. As LLMs increasingly influence developer decisions \cite{decisionmaking}, evaluation outcomes shape both organizational confidence and individual reliance on these systems.

\textbf{Evaluation as Engineering Feedback and Regression Control.} In practice, LLM-based SE tools evolve continuously: prompts, retrieval strategies, guardrails, model versions, and serving configurations are frequently updated. Evaluation, therefore, plays the role of a Quality Assurance (QA) and regression-testing mechanism: it helps tool builders detect behavioral drift, quantify stability across runs, and identify which changes improve developer outcomes versus merely changing style. Without systematic regression evaluation, teams may unknowingly deploy updates that reduce reliability on rare-but-critical cases while appearing ``better'' on superficial metrics.

\textbf{Characterizing Model Behavior Beyond Individual Outputs.}
Unlike traditional software components, LLM-based tools do not expose their behavior through fixed, deterministic outputs. Instead, they operate as general-purpose components whose behavior depends on prompts, context, and usage scenarios \cite{liu2024reliability, ye2024investigating}. Evaluation, therefore, plays a crucial role in characterizing typical model behavior, variability across runs, and sensitivity to context. By moving beyond anecdotal examples or single outputs, evaluation enables a more systematic understanding of how LLM-based tools behave across inputs, tasks, and usage conditions. This characterization is essential for comparing tools, identifying appropriate use cases, and setting realistic expectations about model capabilities.

\textbf{Efficiency, Cost, and Practical Feasibility.}
While accuracy and output quality are important, organizations adopting LLM-based SE tools must also consider efficiency and cost. In real SE workflows, models are expected to provide near-interactive responses, scale across large codebases, and integrate into continuous development pipelines. Evaluation, therefore, plays a critical role in revealing whether a tool can meet these practical requirements. By assessing dimensions such as latency, resource consumption, and monetary cost alongside output quality, evaluation enables informed trade-offs between model capability and deployability. For example, two models may achieve similar accuracy or usefulness, yet differ substantially in inference time or operational cost, leading to very different adoption outcomes in practice. Evaluation that captures these dimensions supports realistic deployment decisions rather than purely performance-driven comparisons.
As LLM-based tools transition from research prototypes to production systems, evaluation becomes a key mechanism for aligning model performance with organizational constraints and expectations. Incorporating efficiency and cost into evaluation is therefore essential for understanding the real-world viability of LLM-based approaches in software engineering.

\textbf{Evaluation for Reliability Awareness.}
Beyond measuring performance, evaluation plays a central role in making the behavior of LLM-based SE tools visible and interpretable. LLMs can generate fluent and persuasive outputs that vary in correctness, support, or appropriateness \cite{jain2025mitigatingCodeLLMHallucinations, ravichander2025halogen}. In software engineering contexts, such variability may not be immediately apparent from individual outputs, yet it can substantially influence developer decisions.

Through systematic evaluation, practitioners and researchers gain insight into the conditions under which LLM-based tools behave reliably and where their limitations lie. Evaluation therefore enables informed use by revealing patterns of behavior that would otherwise remain hidden, supporting realistic expectations about model capabilities and guiding appropriate integration into SE workflows.

%\paragraph{\textbf{Evaluation as an Enabler of Responsible AI4SE}}
%Taken together, these considerations highlight why evaluation occupies a central position in the development and deployment of LLM-based AI4SE systems. Evaluation choices directly affect trust, adoption, comparability, and safety. As LLMs continue to permeate SE practice, careful evaluation becomes necessary not only for measuring performance, but also for enabling responsible, evidence-based use of probabilistic and open-ended systems.

\section{Current State of LLM Evaluation}
\vspace{-1mm}
The evaluation of LLMs has evolved rapidly in response to their growing adoption across diverse tasks and domains. Rather than relying on a single methodology, current practice employs a combination of human judgment, automatic metrics, benchmarks, LLM-based judges, and comprehensive evaluation frameworks. Each of these approaches reflects different assumptions about what constitutes good model behavior and is suited to different types of tasks.
Table~\ref{tab:llm-eval-comparison} summarizes the main evaluation approaches currently used for large language models and highlights their distinct roles within the evaluation pipeline. The following subsections discuss each approach in more detail.
%Rather than ranking or judging these approaches, the table provides a structured overview of the evaluation mechanism, whether reference data is required, the degree of automation involved, and the functional role each approach plays in practice. 

\begin{table*}[t]
\centering
\caption{Overview of current LLM evaluation approaches and their roles in the evaluation pipeline.}
\label{tab:llm-eval-comparison}
\small
\setlength{\tabcolsep}{6pt}
\begin{tabular}{l l l l l}
\toprule
\textbf{Approach} &
\textbf{Evaluation mechanism} &
\textbf{Reference required} &
\textbf{Degree of automation} &
\textbf{Role in the evaluation pipeline} \\
\midrule
Manual evaluation &
Human evaluators &
No &
Manual &
Provides direct human assessment \\

Reference-based metrics &
Automatic metric &
Yes &
Fully automatic &
Quantifies performance against references \\

Benchmarks &
Automatic metric &
Yes &
Fully automatic &
Standardizes comparison across models \\

LLM-as-a-judge &
LLM &
No &
Fully automatic &
Approximates human judgment at scale \\

Evaluation frameworks &
Automatic Metric + LLM &
Optional &
Semi- to fully automatic &
Orchestrates and integrates evaluation methods \\
\bottomrule
\end{tabular}
\end{table*}

\textbf{Manual Evaluation}
Manual evaluation remains a widely used approach for assessing LLM behavior, particularly for tasks with open-ended or subjective outputs \cite{chang2024survey}. In this setting, human evaluators assess model responses directly according to qualitative criteria defined by the evaluation protocol. This form of evaluation can be employed for tasks such as code review, explanation, summarization, and conversational interaction, where automatic metrics struggle to capture nuanced aspects of model behavior \cite{chang2024survey}.

Manual evaluation is often used as a reference point for assessing model performance and for validating automated evaluation methods. It is especially prevalent in scenarios where multiple responses may be acceptable and where correctness cannot be determined through fixed ground truth alone \cite{wagner2025towards}. As a result, manual evaluation continues to play a central role in the evaluation of LLM-based systems despite the emergence of increasingly automated alternatives.

\textbf{Reference-Based Automatic Evaluation.}
Reference-based automatic evaluation assesses model outputs by comparing them against predefined gold references. This approach underlies many traditional evaluation metrics, including accuracy, BLEU \cite{papineni2002bleu}, ROUGE \cite{lin2004rouge}, and METEOR \cite{banerjee2005meteor} has been widely adopted in tasks such as classification, translation, summarization, and code generation \cite{zhang2023survey, lu2021codexglue}. When clear ground truth exists, reference-based evaluation provides a straightforward and reproducible way to quantify model performance.
Within LLM evaluation, reference-based metrics remain commonly used for tasks that can be expressed as well-defined input–output mappings. These metrics are frequently employed in benchmarks and experimental studies where standardized evaluation is required and where reference outputs are available.

\textbf{Benchmarks.}
Benchmarks provide standardized datasets, tasks, and evaluation protocols that enable systematic comparison across models. The recent growth of LLM research has led to the development of numerous benchmarks targeting different capabilities, including reasoning \cite{sawada2023arb}, factual and domain knowledge \cite{hendrycks2020measuring}, and SE tasks \cite{jimenez2023swe}. Examples include benchmarks designed to evaluate realistic programming scenarios like SWE-Bench as well as those focusing on broader cognitive or problem-solving abilities.

Benchmark-based evaluation plays an important role in tracking progress over time and facilitating comparison across models and approaches. By fixing datasets and evaluation procedures, benchmarks offer a common reference point for reporting and interpreting results within the research community.

\textbf{LLM-as-a-Judge.}
LLM-as-a-judge approaches use an LLM to evaluate the outputs of another model. In this paradigm, an LLM is prompted to assess responses based on task-specific instructions or rubrics, often producing scores, rankings, or preference judgments. This approach has gained traction as a flexible and scalable way to evaluate tasks where reference answers are unavailable or difficult to define \cite{wang2025canllms, zheng2023judging}.

In current practice, LLM-as-a-judge methods are applied across a range of tasks, including dialogue evaluation \cite{feng2025samer}, reasoning assessment \cite{lee-etal-2025-checkeval}, qualitative comparison of model outputs \cite{kim2025aura}, and also in the SE context \cite{wang2025canllms}. They are often used to approximate human judgments or to support large-scale evaluations where manual assessment would be impractical \cite{zheng2023judging}.

%In practice, this separation may involve using one LLM to generate artifacts (e.g., code or reviews) and a different LLM to evaluate them, reflecting the decoupling of generation and evaluation roles in real-world pipelines.

\textbf{Evaluation Frameworks.}
Evaluation frameworks aim to provide structured and reusable pipelines for conducting LLM evaluation. Frameworks such as DeepEval, RAGAS, HELM, %Langchain\footnote{https://www.langchain.com/langsmith/evaluation} 
offer abstractions for defining evaluation criteria, running experiments, and aggregating results across models and tasks.
By encapsulating evaluation logic into configurable components, these frameworks support repeatable experimentation and reduce the engineering effort required to evaluate LLM-based systems. They are increasingly used to operationalize evaluation practices across research and applied workflows.
Altogether, the current state of LLM evaluation is characterized by methodological diversity. Manual evaluation, reference-based metrics, benchmarks, LLM-as-a-judge approaches, and evaluation frameworks are all actively used, often in combination. Rather than converging on a single standard, evaluation practice reflects the heterogeneity of LLM tasks and application contexts, particularly in SE settings.

Table~\ref{tab:task-characteristics} summarizes how representative SE tasks differ in their evaluation characteristics, highlighting the role of task objectives and output structure in shaping evaluation difficulty.

\begin{table*}[t]
\centering
\caption{Software engineering tasks and their evaluation characteristics.}
\label{tab:task-characteristics}
\footnotesize
\setlength{\tabcolsep}{5pt}
\begin{tabularx}{\textwidth}{l Y Y Y}
\toprule
\textbf{SE task} &
\textbf{Task description} &
\textbf{Output characteristics} &
\textbf{Why is evaluation easy or problematic?} \\
\midrule

Code generation \cite{chen2024survey, zhang2023planning}&
Generate executable code from specifications or prompts &
Structured, executable code &
Correctness can be checked via unit tests or execution outcomes \\
\addlinespace

Bug classification / triage \cite{aracena2024applying, kumar2024ensemble}&
Assign bugs to categories such as type, severity, or priority &
Discrete labels from predefined classes &
Explicit labels provide a clear (though imperfect) ground truth \\
\addlinespace

Code review \cite{cihan2025automated, lu2023llama}&
Provide feedback on code changes or pull requests (PR) &
Open-ended natural language comments &
Quality depends on usefulness and context; no single correct answer \\
\addlinespace

Code explanation \cite{macneil2022generating, bhattacharya2023exploring} &
Explain code behavior or design decisions &
Open-ended natural language explanations &
Plausible but shallow explanations may appear correct \\
\addlinespace

Program repair \cite{huang2025comprehensive, li2025hybrid} &
Generate patches that fix faulty code &
Structured code with executable semantics &
Correctness can be verified by tests, but multiple fixes may exist \\
\addlinespace

Effort/cost/time estimation \cite{bui2025llm, calikli2025request-format-effort-estimation} &
Estimate development effort, time, or cost for a task or change &
Numeric or ordinal estimates &
Estimates are inherently subjective and influenced by context and uncertainty \\

\bottomrule
\end{tabularx}
\end{table*}

\section{Challenges of Current LLM Evaluation}

Consider an LLM-based automated code review assistant that comments on PRs. If the evaluation protocol uses a reference comment set and a lexical overlap metric (or a single ``useful/not useful'' judge prompt), the system may score highly by producing fluent, stylistically similar feedback while systematically missing rare-but-critical issues (e.g., input validation gaps, unsafe error handling, or security-relevant edge cases). In practice, developers may still accept the tool’s suggestions because they read plausibly, yet the evaluation signal fails to reflect the risk profile that matters for deployment. This example illustrates a broader pattern in current LLM evaluation: evaluation signals that appear reasonable in isolation can fail to capture the risks and behaviors that matter in real-world software engineering contexts.%This illustrates why AI4SE evaluation must (i) make the evaluation target explicit (what “good” means for developers), (ii) account for non-determinism via multi-run uncertainty, and (iii) incorporate verification and guardrails (tests/static analysis/abstention) while reporting the resulting latency and cost overhead.

Despite the rapid development of evaluation tools and practices, the evaluation of LLM-based systems continues to face fundamental challenges that limit the reliability, interpretability, and usefulness of evaluation results. These challenges arise not from a single source, but from the interaction between non-deterministic model behavior, open-ended outputs, human judgment, and the complexity of real-world SE contexts. In the following, we discuss these challenges briefly.

\textbf{Absence of Stable and Objective Ground Truth.}
A central challenge in LLM evaluation is the lack of stable and objective ground truth for many tasks \cite{jiang2024genresrethinkingevaluationgenerative}. While reference answers can be defined for closed-form problems, a large fraction of LLM-driven AI4SE tasks, such as code review, code explanation, bug analysis, and requirements interpretation, do not admit a single correct answer. Instead, multiple responses may be acceptable depending on context, assumptions, and developer intent.

Even when human annotations are used to construct reference labels, treating these labels as ``gold'' introduces additional uncertainty. Prior works \cite{dougan2019investigating, tuzun2022ground} have shown that ground truth derived from historical human decisions may be incomplete, biased, or misaligned with the intended evaluation objective. For example, studies on ground truth deficiencies \cite{tuzun2022ground} in SE report that recorded labels often reflect convenience, availability, or organizational constraints rather than optimal technical decisions, which can distort evaluation outcomes. Similarly, research on code reviewer recommendation \cite{dougan2019investigating} demonstrates that reviewers assigned in practice are frequently treated as ideal ground truth despite evidence that assignments are influenced by non-technical factors such as workload and social dynamics.

Insights from cognitive psychology further reinforce these concerns. As argued by Kahneman in \textit{``Thinking, Fast and Slow''} \cite{kahneman_thinking_2012}, human evaluators rely on fast, intuitive reasoning that is sensitive to framing, prior beliefs, and contextual cues. More directly, \textit{``Noise: A Flaw in Human Judgment''} \cite{kahneman2021noise} shows that even trained experts can produce substantially different judgments when evaluating identical cases, a phenomenon referred to as judgment noise.
Taken together, these findings challenge the assumption that human-labeled data constitutes a reliable gold standard for evaluation. In the context of LLM evaluation, disagreement, noise, and bias in human judgments can propagate into benchmarks and evaluation protocols, undermining reproducibility and comparability across studies. This issue is particularly pronounced in evaluation setups that rely on a small number of annotators or do not explicitly model inter-annotator disagreement.

\textbf{Subjectivity and Multi-Dimensional Quality.}
Many existing evaluation approaches discussed above implicitly assume that model quality can be reduced to a single scalar score. However, for open-ended LLM outputs, quality is inherently multi-dimensional. Outputs may differ in usefulness, completeness, reasoning style, or alignment with developer expectations, even when they are all technically plausible.

As highlighted in recent surveys \cite{chang2024survey, mahaut-etal-2024-factual} of LLM evaluation, reducing such outputs to a single metric often obscures important qualitative differences and can lead to misleading comparisons. This challenge is compounded in SE tasks, where judgments depend not only on textual correctness but also on implicit domain knowledge, project conventions, and downstream impact on developer decisions.

\textbf{Non-Determinism and Evaluation Instability.}
LLMs are inherently stochastic systems. Due to probabilistic decoding and sensitivity to prompt phrasing, repeated runs of the same model under identical conditions can produce different outputs. Crucially, even when the temperature is set to zero, system-level non-determinism in high-performance environments prevents these models from behaving as truly deterministic functions \cite{ouyang2025empirical, atil-etal-2025-non}. This non-determinism complicates evaluation in two ways.

First, single-run evaluations may fail to represent typical model behavior. Second, evaluation results themselves can vary across runs, even when using automatic metrics or LLM-based judges. The inherent variability in LLM performance—driven by factors such as training data, architecture, and inconsistent evaluation methods—often results in irreproducible findings. Consequently, poorly controlled evaluation setups undermine reproducibility and make fair comparison between models difficult \cite{wagner2025towards}.
In SE contexts, where evaluation results may influence deployment or adoption decisions, such instability poses a serious risk. Without explicit strategies for repeated evaluation, aggregation, or uncertainty reporting, evaluation outcomes can give a false sense of confidence.

\textbf{Limitations of Automated and Model-Based Evaluation.}
Automated evaluation methods, including reference-based metrics and LLM-as-a-judge approaches, offer scalability but introduce their own challenges. Reference-based metrics rely on surface-level similarity and assume that reference outputs adequately capture task intent, an assumption that often fails for open-ended AI4SE tasks. LLM-as-a-judge methods, while flexible, inherit biases, blind spots, and inconsistencies from the models they employ.

Empirical studies \cite{ye2024justice, shi-etal-2025-judging} have shown that LLM-based judges do not always align with human preferences and may exhibit systematic biases depending on prompt design, model choice, or evaluation context. This raises concerns about circularity, where models are effectively evaluated by systems with similar training data and inductive biases.

%\paragraph{\textbf{Mismatch with SE Practice}}
%Finally, many current evaluation practices operate at a level of abstraction that does not fully reflect how LLM-based tools are used in SE practice. Evaluations often focus on isolated inputs and outputs, whereas real-world usage involves interaction, iteration, and integration into broader workflows. As a result, evaluation results may fail to capture how models influence developer decisions over time or how errors propagate in realistic settings.

\textbf{Fragmentation of Evaluation Practices.}
Current LLM evaluation practice is highly fragmented, with different tools, metrics, and frameworks emphasizing distinct aspects of model behavior \cite{jiang2024genresrethinkingevaluationgenerative}. As a result, the same system may appear strong or weak depending on the chosen evaluation setup. This fragmentation complicates comparison across studies and makes it difficult to accumulate reliable empirical knowledge. In SE contexts, where evaluation results may guide tool adoption or deployment, inconsistent evaluation signals can lead to conflicting conclusions and undermine trust in reported results.

Together, these challenges indicate that LLM evaluation is not merely a technical measurement problem, but a broader socio-technical issue involving human judgment, probabilistic systems, and SE practice.

\section{Open Research Questions and Directions}
\vspace{-1mm}
%The challenges outlined in the previous section point to a number of open research questions that remain largely unanswered in current LLM evaluation practice. 
%From a vision perspective, we argue that evaluation in AI4SE should be treated as an SE-grade quality loop: (i) explicitly specify task intent, context scope, and risk tolerance; (ii) characterize behavior under non-determinism using multi-run protocols and uncertainty reporting; (iii) combine executable checks, calibrated automated judging, and targeted human audits to assess multi-dimensional quality; and (iv) report not only quality signals (e.g., accuracy or usefulness) but also operational trade-offs (latency, resource use, and monetary cost) that determine deployability.

We see four near-term standardization targets for trustworthy AI4SE evaluation:
(1) lightweight task specifications (“evaluation cards”) that encode intent, context scope, and risk tolerance;
(2) default multi-run reporting (distributions + rank stability) for both generators and judges;
(3) developer-centric, multi-dimensional rubrics with severity weighting (e.g., security-critical misses dominate style); and
(4) calibrated hybrid judging pipelines that combine executable checks (tests/static analysis), LLM judges, and small human audit sets.
%Across all four, evaluation should report quality–cost–latency trade-offs and treat guardrails as first-class components of the evaluated system.

\textbf{Evaluating open-ended tasks without stable ground truth.}
A fundamental open question concerns how open-ended LLM tasks should be evaluated
in the absence of a single, objective ground truth. Many AI4SE tasks admit multiple valid
responses, making traditional reference-based evaluation insufficient. This raises questions
about whether evaluation should move beyond correctness-oriented judgments toward
comparative, preference-based, or outcome-oriented assessment, and how such approaches
can be designed to remain systematic, reproducible, and interpretable.
A promising direction is to make the evaluation target explicit via lightweight evaluation
specifications (e.g., intended developer goal, acceptable variation, severity weighting) so that
evaluation rewards “useful and risk-aware” outputs rather than merely “similar” ones.

\textbf{Defining and detecting hallucinations across tasks.}
%A closely related challenge is the reliable detection of hallucinations, particularly outside RAG settings where explicit evidence is available.
While hallucinations are widely recognized as a critical failure mode, there is still no consensus on how
to define, operationalize, or measure them across different tasks. Open questions remain regarding
whether hallucination detection should rely on external signals, internal model uncertainty, cross-model
agreement, or human-in-the-loop verification, and how these signals should be combined.
In AI4SE settings, an important pathway forward is to couple hallucination evaluation with
tool-level guardrails, and to quantify both their risk-reduction
benefits and their latency/cost overhead.

\textbf{Accounting for non-determinism and evaluation variability.}
Non-determinism introduces another set of open problems for evaluation. Given that repeated runs of
the same model can yield different outputs, it remains unclear how evaluation protocols should account
for variability. Key questions include how many runs are sufficient to characterize typical model behavior,
how evaluation results should be aggregated, and how uncertainty should be reported to support fair
comparison and reproducibility.
A practical direction is to standardize multi-run evaluation as a default reporting practice and to treat
key outcomes as distributions (including rank stability), while explicitly budgeting the added evaluation cost and runtime that multi-run protocols introduce.

\textbf{Integrating human judgment without treating it as ground truth.}
Human evaluation, despite its central role, raises unresolved questions about how human judgment should
be incorporated into evaluation pipelines in a principled way. Well-documented noise and inconsistency
in human decisions challenge the assumption that human labels constitute reliable ground truth. Open
questions include how many annotators are required, how disagreement should be modeled rather than
eliminated, and how human judgments can be combined with automated signals without masking uncertainty.
One promising approach is to treat humans as calibration and audit signals (with explicit disagreement
modeling) and to use automated methods for scale only after they are validated against small, high-quality
audit sets.

\textbf{Evaluating guardrails and deployment-time controls.}
In practice, LLM-based SE tools are rarely deployed “naked”; they are wrapped with guardrails such as
constrained prompting, retrieval grounding, policy filters, output validators, sandboxed execution, and
fallback/abstention rules. An open question is how to evaluate guardrails as first-class components rather
than incidental implementation details: how should we measure their marginal contribution to reliability and
risk reduction, and how do we choose guardrail configurations that satisfy latency and cost budgets?
Answering this requires evaluation protocols that report end-to-end tool performance (model + guardrails),
including both quality outcomes and the operational overhead introduced by safety mechanisms.

\textbf{Balancing evaluation rigor with practical constraints.}
Finally, there is an open tension between academic evaluation rigor and industrial practicality. Many evaluation
protocols are expensive, time-consuming, or difficult to integrate into real-world development workflows.
At the same time, overly simplified evaluation risks missing important weaknesses. A key research question is
how evaluation methods can balance methodological rigor with scalability, cost, and relevance to practitioners.
A useful framing is to treat evaluation itself as a resource-bounded process (time, money, and human attention),
and to develop tiered evaluation pipelines (cheap automated checks $\rightarrow$ calibrated judging $\rightarrow$
targeted human audits $\rightarrow$ optional workflow-outcome validation) that exposes explicit quality--cost--latency
trade-offs.

\bibliographystyle{ACM-Reference-Format}
\bibliography{references}

\end{document}